# Mileage-responsive Wind Power Smoothing

Xue Lyu*, *Student Member, IEEE*, Youwei Jia*, *Member, IEEE*, Zhao Xu, *Senior Member, IEEE*
Jacob Østergaard, S*enior Member, IEEE*

*Abstract*-- **This paper proposes a novel wind power smoothing control paradigm in context of performance-based regulation service. Conventional methods aim at adjusting wind power output using hard-coded filtering algorithms that can result in visually smoothed power output with unmeasurable impacts on system generation-demand balance. Distinguished from conventional methods, the newly proposed control method smooths wind power output from a power system perspective by using the regulation mileage as a key performance indicator. To simultaneously address the system needs and maximize wind energy harvesting, a mileage-responsive framework is developed to enable wind farms to optimally generate smoothing power. The effectiveness of the proposed method is well demonstrated through case studies, of which the simulation results shows a great potential for practical applications.**

*Index Terms*--wind energy, smoothing control, mileage, performance-based regulation service

## I. Introduction

TILL date, various wind power smoothing approaches have been reported [1] whereas a general scientific question—to which extent such "smoothed" wind power could indeed contribute to the grid operation—still remains open. In this context, this paper particularly holds the perspective that wind power smoothing control shall be intrinsically correlated with the main grid operation.

Admittedly, grid frequency stability can be significantly crippled by instant generation-demand imbalance. Smoothing control is aimed to alleviate power fluctuations caused by intermittent wind speed, so that the resulting system power imbalance can be more or less mitigated instantaneously. In the literature, some pioneering work on power smoothing might be less effective due to the following reasons: 1) the control reference of power output is typically generated by hard-coded algorithms (e.g. moving average[2], low-pass filter[3], ramp limits[4]). Such references are only dependent on autogenous wind power generation profiles yet are weakly correlated with grid frequency stabilization; 2) there lacks a consideration of wind turbine (WT) operational constraints, such that obtained references can be technically ineffective in terms of unreasonable over-/under-production demand; and 3) it can be economically inefficient by arbitrarily sacrificing a significant amount of wind energy through predefined hard-coded smoothing mechanism.

This paper is dedicated to addressing the aforementioned drawbacks by incorporating wind power smoothing control with its consequent effects on system frequency regulation service cost. Towards this end, we newly develop a receding horizon control framework, which consists of an optimization module to generate optimal control reference by considering wind-fluctuation-induced frequency stabilization cost, and a real-time control module to timely track the smoothing reference. It is worth noting that the proposed framework is distinguished among most existing works in this field and holds significant merits as 1) smoothing reference is obtained through system-perceived optimization; and 2) the effectiveness of WT over-/under-production at instant moments can always be ensured through practical-constrained cascade control.

## II. Problem Description

In performance-based regulation market, automatic generation control (AGC) units are engaged to provide frequency regulations with specific service charges, which covers regulation capacity payment and regulation mileage payment [5]. From the perspective of service providers, mileage payment indicates the degree of frequency regulation engagement for a certain period. Likewise, it is reasonable to extend this concept to WTs since their non-predictable power fluctuations lead to such frequency regulation burden of AGC units. Analogous to mileage payment, we refer such fluctuation-induced service charge as mileage cost at WT point of view. In such sense, mileage cost can quantitively reflect the degree of wind power fluctuations, of which the magnitude is strongly correlated with AGC engagement.

From WT point of view, we formulate the smoothing task as a cost-oriented optimization problem which is tackled by model predictive control. In particular, this problem can be described on two different levels. For the system optimization level, optimal control references (i.e. wind power command) should simultaneously consider the minimum induced mileage cost and maximum wind energy harvesting in a specified look-ahead time window. Conducting sufficiently long look-ahead optimization can undoubtedly ensure the obtained power demand being close to the global optima. For the WT control level, following the

This work is partially supported by Hong Kong RGC Theme Based Research Scheme Grants No. T23-407/13N and T23-701/14N.
X. Lyu, Y. Jia, Z. Xu are with Department of Electrical Engineering, The Hong Kong Polytechnic University, Hung Hom, Hong Kong. (E-mails: xue.lyu@connect.polyu.hk,corey.jia@connect.polyu.hk, eezhaoxu@polyu.edu.hk)
J. Østergaard is with the Center for Electric and Energy, Department of Electrical Engineering, Technical University of Denmark, Kongens Lyngby 2800, Denmark. (E-mail: joe@elektro.dtu.dk)
* The authors are co-first authors

optimized control reference in real time necessitates over-/under-production of WTs to a satisfactory extent.

Obviously, the proposed smoothing task in our paper is fundamentally new while technically challenging to handle. On the one hand, there is a prerequisite to model power system AGC considering system control architecture and operation scenarios so that the corresponding WT mileage cost can be quantified. On the other hand, WT modeling contains high non-linearity, complex formulation can make the optimization process intractable. In the forthcoming section, the proposed methodology will be discussed in details.

## III. METHODOLOGY

To realize the proposed system-accepted wind power smoothing task, a model predictive control framework is developed (as shown in Fig.1). In essence, the overall smoothing performance relies on the synergy between optimization module and WT control module.

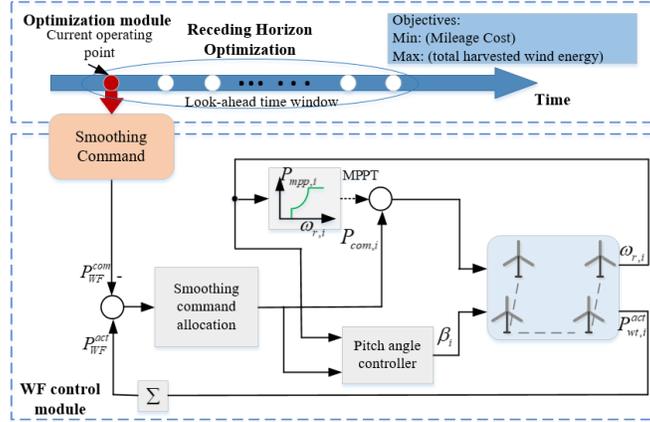

Fig.1 The proposed mileage-inspired wind power smoothing control framework

### A. Receding Horizon Optimization

In a specified look-ahead time window (i.e. from present $t_0$ to future time instance $t_T$), the smoothing task is formulated as the following optimization problem.

$$\min \ \alpha \int_{t_0}^{t_T} -(\sum_{i=1}^{N} P_{Ei}^t) dt / E_{base}^{wf} + (1-\alpha) \int_{t_0}^{t_T} (\sum_{i=1}^{n} (\gamma_t g_t^i)^2) dt / (C_{base}^{mileage})^2 \quad (1)$$

where $\gamma_t$ is mileage-based clearing price in regulation markets at time $t$, $g_t^i$ represents the regulation power of AGC unit $i$ at time $t$, $E_{base}^{wf}$ and $C_{base}^{mileage}$ are base values for normalizing energy harvesting of the wind farm and mileage cost, respectively. $P_{Ei}^t$ is the electrical power output of WT $i$. Based on the swing equation of the turbine, $P_{Ei}^t$ can be calculated using the following equation,

$$P_{Ei}^t = 0.5 \rho A v_{ti}^3 C_{Pi}^t - J\omega_i(t)[\omega_i(t) - \omega_i(t-1)] / \Delta t \quad (2)$$

where $\rho$ is air density, $A$ is rotor swept area, $v_t$ is wind speed, $C_P$ is power coefficient, $J$ is the equivalent moment of inertia of turbine blades and generator.

The objective function above integrates two independent normalized objectives (i.e. maximizing wind energy harvesting and minimizing wind fluctuation induced mileage cost) with weighting factors $\alpha$ and $1-\alpha$, where $\alpha \in [0,1]$. It should be noted that assigning different $\alpha$ results in different trade-offs in the Pareto front. In practice, $\alpha$ can be determined by system operators in accordance with different grid operational status. For example, assigning a small value of $\alpha$ can release the pressure of AGC units to some extent to counteract grid frequency instability. On the contrary, a wind farm is supposed to generate as much power as possible in normal operation status. In such circumstance, a large $\alpha$ is desired.

In quantifying the mileage cost, the clearing price $\gamma_t$ is typically determined by ISO based on specific market information, which covers bid-in regulation capacity, mileage price, and historical performance indices of AGC units [5]. Detailed price-clearing mechanism is essential yet beyond the scope of the current work, hence will not be discussed further in this paper. Interested readers can refer to many literatures available (e.g. [5] ).

In modelling the total amount of wind energy harvesting, $C_P$ normally involves high non-linearities with WT rotor speed and pitch angle. For simplicity, we adopt the following polynomial regression, which gives rise to satisfactory modeling accuracy.

$$C_{Pi}^t = [c_{11}\beta_i(t)^2 + c_{12}\beta_i(t) + c_{13}]\lambda_{ti}^2 + [c_{21}\beta_i(t)^2 + c_{22}\beta_i(t) + c_{23}]\lambda_{ti} + [c_{31}\beta_i(t)^2 + c_{32}\beta_i(t) + c_{33}] \quad (3)$$

where $\beta$ is pitch angle; $c_{ij}$ is regression parameter; and $\lambda_t$ is tip speed ratio.

The electric output power of WTs is subject to the following constraints,

$$0 \leq P_{Ei}^t \leq P_{MPPi}^t + P_{Ki} \quad (4)$$

where $P_{MPP}^t$ is the active power reference generated through the maximum power point tracking (MPPT) algorithm. $P_K$ is the maximum electric power that can be converted from the kinetic energy stored in the rotational rotor, which is mathematically expressed as,

$$P_{Ki} = 0.5J[\omega_i(t-1)^2 - \omega_{MPPi}(t)^2]/\Delta t \quad (5)$$

where $\omega$ denotes the WT rotor speed; $\omega_{mpp}$ is the rotor speed in MPPT operation mode.

WTs operation should also be subject to the following practical constraints.

$$\omega_{min} \leq \omega_i(t) \leq \omega_{max} \quad (6)$$

$$\beta_{min} \leq \beta_i(t) \leq \beta_{max} \quad (7)$$

At the system level, AGC should be subject to the following power balance constraints.

$$P_{scheduled}^t + P_E^t + \sum_{i=1}^n g_t^i = P_{load}^t \quad (8)$$

$$-g_{cap}^i \leq g_t^i \leq g_{cap}^i \quad (9)$$

where $P_{scheduled}^t$ is the scheduled power generation in the power grid at time $t$ corresponding to the instantaneous system load demand $P_{load}^t$, and $g_{cap}^i$ is the regulation capacity of AGC unit $i$.

*B. Wind Turbines Cascade Control*

Through receding horizon optimization, WTs output power needs to be controlled to track the optimized power command in real-time. In this paper, a cascade strategy is proposed to realize optimal smoothing control. As illustrated in Fig.2, the proposed control strategy is not dependent on accessional energy storage system. It takes full advantage of the WT self-capability by sequentially adopting rotor speed control and pitching control to fulfill the command.

Echoed with the first optimization objective of maximizing wind energy harvesting, rotor speed control would preferentially be utilized to achieve over-/under-production by releasing/storing excessive kinetic energy from the rotating rotor. This part of controller features high energy utilization efficiency in managing redundant or insufficient wind energy through WTs. Nevertheless, the active power support provided via rotor speed control is subject to strict WT physical constraints as expressed in (5). As shown in Fig.2, in case that rotational speed reaches its upper limit (i.e. $\omega \geq \omega_{max}$), pitching control will be activated to offer further power curtailment capability. Pitching is at the lowest priority in the proposed strategy so as to reduce the wastage of wind energy and mitigate the fatigue of WTs.

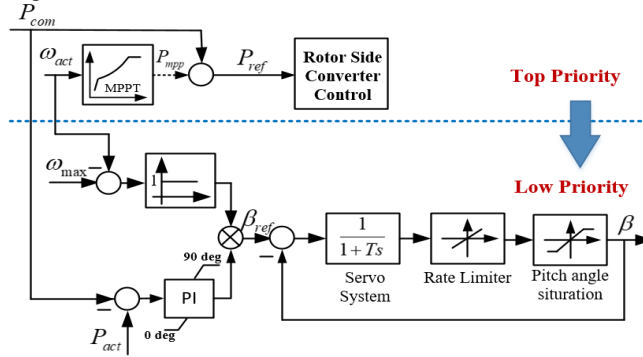

Fig.2 Proposed cascade control for wind power smoothing

## IV. CASE STUDIES

The proposed method is preliminarily tested on a simple-but-typical operation scenario, which is illustrated in Fig.3. In this scenario, the instantaneous power imbalance (i.e. $P_{net-load}$-$P_{scheduled}$) is complemented by three AGC units. Hence, it is straightforward that reduction of wind power fluctuations can directly reduce AGC mileage cost. Some simulation parameters are predefined in Table I. The price clearing mechanism can be referred to [5].

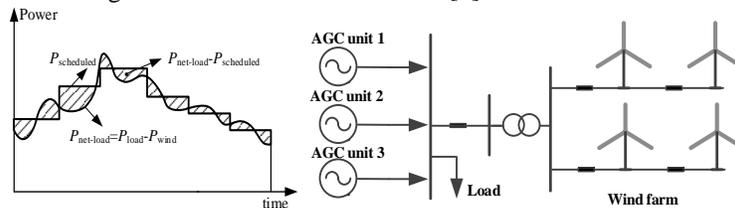

Fig. 3 Testing operation scenario

Table I Simulation parameters

| AGC | $p_m$: normalized regulation performance score<br>$P^i_{offer}$: mileage bidding price of unit $i$ ($/MW)<br>$g^i_{cap}$: maximum biddable capacity of unit $i$ (MW) |
|---|---|
| | $p^1_m = 0.7168 \quad P^1_{offer} = 2\$/MW \quad g^1_{cap} = 1.5$ |
| | $p^2_m = 0.6074 \quad P^2_{offer} = 4\$/MW \quad g^2_{cap} = 4$ |
| | $p^3_m = 1 \quad P^3_{offer} = 1\$/MW \quad g^3_{cap} = 2.5$ |
| Wind speed data | Online accessible [6] |
| Wind farm | The studied wind farm consists of 4 homogeneous wind turbines with total 20 MW generation capacity. |

In this case study, a typical time series of wind speed data for 1 hour is considered. The formulated non-linear optimization problem is solved every 4 seconds to update the power command, which is consistent with the AGC control cycle. In our study, the optimization part can be readily solved in 0.2732s (this computation time is averaged by 10000 individual simulations based on Dell Precision Tower with CPU E5-2650 v4@2.20GHz (2 processors)).

It should be noted that the overall smoothing performance can be influenced by different weights applied to wind energy harvesting and mileage cost. Fig. 4 shows the simulation results given that $\alpha = 0.3$ and $\alpha = 0.7$, respectively. The detailed results on total wind energy generation and induced mileage cost are reported in Table II. Obviously, free-running wind farm based on MPPT control generally produces the most fluctuated power which requires large power balancing effort from the main grid (i.e. considerable mileage cost). In contrast, the proposed method can provide measurable and adjustable benefits to the main grid by tuning the tradeoff between wind energy harvesting and the induced AGC mileage cost. Such Pareto decision-making is dependent on the actual system needs.

Table II Simulation results (for 1 hour) obtained by different methods

| | | Wind energy production (kWh) | Mileage cost ($) |
|---|---|---|---|
| Proposed method | $\alpha=0.3$ | $8.128\times10^3$ | $2.61\times10^3$ |
| | $\alpha=0.7$ | $8.221\times10^3$ | $2.68\times10^3$ |
| MPPT | | $9.6817\times10^3$ | $1.1708\times10^4$ |

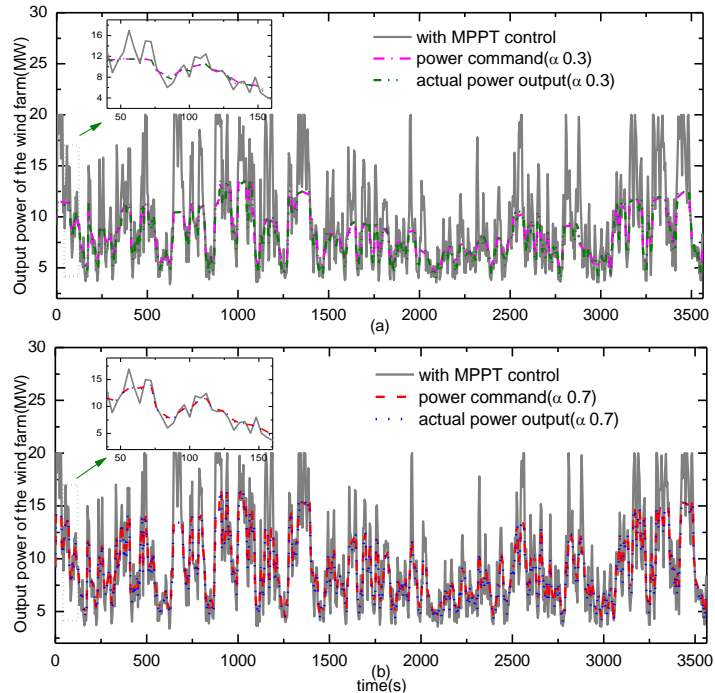

Fig.4 Smoothing results via the proposed strategy (a. $\alpha = 0.3$, b. $\alpha = 0.7$)

## V. CONCLUSION

This paper newly proposes an optimal wind power smoothing control framework, which is successfully tested through a preliminary case study. It is highlighted that the proposed system-oriented smoothing control is advantageous over conventional hard-coded filtering based algorithms to ensure genuinely smoothed operation of WTs subject to various constraints. Future work is underway to further enhance the proposed smoothing framework by taking the wake interactions among turbines into account.

## REFERENCES

[1] A. M. Howlader, N. Urasaki, A. Yona *et al.*, "A review of output power smoothing methods for wind energy conversion systems," *Renewable and Sustainable Energy Reviews,* vol. 26, pp. 135-146, 2013.

[2] M. Chowdhury, N. Hosseinzadeh, and W. Shen, "Smoothing wind power fluctuations by fuzzy logic pitch angle controller," *Renewable Energy,* vol. 38, no. 1, pp. 224-233, 2012.

[3] A. Uehara, A. Pratap, T. Goya *et al.*, "A coordinated control method to smooth wind power fluctuations of a PMSG-based WECS," *IEEE Transactions on energy conversion,* vol. 26, no. 2, pp. 550-558, 2011.

[4] Q. Jiang, and H. Hong, "Wavelet-based capacity configuration and coordinated control of hybrid energy storage system for smoothing out wind power fluctuations," *IEEE Transactions on Power Systems,* vol. 28, no. 2, pp. 1363-1372, 2013.

[5] A. Sadeghi-Mobarakeh, and H. Mohsenian-Rad, "Optimal bidding in performance-based regulation markets: An mpec analysis with system dynamics," *IEEE Transactions on Power Systems,* vol. 32, no. 2, pp. 1282-1292, 2017.

[6] Online available: http://www.winddata.com/